\pdfoutput=1

\documentclass[11pt]{article}

\usepackage{acl}

\usepackage{times}
\usepackage{latexsym}

\usepackage{microtype}
\usepackage{graphicx}
\usepackage{adjustbox}
\usepackage{multirow}
\usepackage{paralist}
\usepackage[inline]{enumitem}
\usepackage{graphicx}
\usepackage{url}
\usepackage{textcomp}
\usepackage{amssymb}
\usepackage{caption}
\usepackage{subcaption}
\usepackage{color,soul}
\usepackage{paralist}
\usepackage{float}
\usepackage{placeins}
\usepackage{aliascnt}
\usepackage{mathtools}
\usepackage{amsmath}
\usepackage{hyperref}

\usepackage[T1]{fontenc}

\usepackage[utf8]{inputenc}

\usepackage{microtype}

\title{Typo-Robust Representation Learning for Dense Retrieval}

\author{
Panuthep Tasawong\textsuperscript{\dag}, Wuttikorn Ponwitayarat\textsuperscript{\dag}, Peerat Limkonchotiwat\textsuperscript{\dag} , \\
\textbf{Can Udomcharoenchaikit}\textsuperscript{\dag}, \textbf{Ekapol Chuangsuwanich}\textsuperscript{‡}, \textbf{Sarana Nutanong}\textsuperscript{\dag}\\
  \textsuperscript{\dag}School of Information Science and Technology, VISTEC, Thailand\\
  \textsuperscript{‡}Department of Computer Engineering,  Chulalongkorn University, Thailand \\
  \texttt{\{panuthep.t\_s20,wuttikorn.p\_s22,peerat.l\_s19}
  \\
  \texttt{,canu\_pro,snutanon\}@vistec.ac.th,}\\
    \texttt{ekapolc@cp.eng.chula.ac.th}
  }

\begin{document}
\maketitle
\begin{abstract}
Dense retrieval is a basic building block of information retrieval applications.
One of the main challenges of dense retrieval in real-world settings is the handling of queries containing misspelled words. 
A popular approach for handling misspelled queries is minimizing the representations discrepancy between misspelled queries and their pristine ones. 
Unlike the existing approaches, which only focus on the alignment between misspelled and pristine queries, our method also improves the contrast between each misspelled query and its surrounding queries.
To assess the effectiveness of our proposed method, we compare it against the existing competitors using two benchmark datasets and two base encoders.
Our method outperforms the competitors in all cases with misspelled queries.
Our code and models are available at \url{https://github.com/panuthept/DST-DenseRetrieval}.
\end{abstract}

\section{Introduction}
Dense retrieval is a fundamental component in many information retrieval applications, such as open-domain question answering and ad-hoc retrieval. 
The objective is to score and rank a large collection of candidate passages based on their similarity to a given query. 
The performance of dense retrieval relies on representation learning.
A popular approach is to finetune a pre-trained language model to create an embedding space that puts each query closer to its corresponding passages~\cite{DBLP:journals/corr/abs-2006-15498, 10.1145/3397271.3401075, xiong2021approximate, qu-etal-2021-rocketqa, ren-etal-2021-pair, ren-etal-2021-rocketqav2}.

One of the major challenges of dense retrieval is the handling of misspelled queries which induces representations of the misspelled queries to be closer to irrelevant passages than their corresponding passages.
Several studies have demonstrated that misspellings in search queries can substantially degrade retrieval performance~\cite{zhuang-zuccon-2021-dealing, 10.1007/978-3-030-99736-6_27}, specifically when informative terms, such as entity mentions, are misspelled~\cite{Sidiropoulos_2022}. 

To create a retrieval model that is capable of handling misspelled queries, researchers have proposed different training methods to align representations of misspelled queries with their pristine ones.
\citet{zhuang-zuccon-2021-dealing, 10.1145/3477495.3531951} devise augmentation methods to generate misspelled queries and propose training methods, Typos-aware Training and Self-Teaching (ST), to encourage consistency between outputs of misspelled queries and their non-misspelled counterparts.
Alternatively, \citet{Sidiropoulos_2022} apply contrastive loss to enforce representations of misspelled queries to be closer to their corresponding non-misspelled queries.
Although these methods can improve the performance of retrieval models for misspelled queries, there is still a substantial performance drop for misspelled queries.

In this paper, we propose a training method to improve dense retrieval for handling misspelled queries based on the following desired properties:
\begin{compactitem}[•]
    \item \textit{\textbf{Alignment}}: the method should be able to align  queries with their corresponding passages.
    \item \textit{\textbf{Robustness}}: the method should be able to align misspelled queries with their pristine queries.
    \item \textit{\textbf{Contrast}}: the method should be able to separate queries that refer to different passages and passages that correspond to different queries.
\end{compactitem}
In contrast to the existing methods for handling misspelled queries that only satisfy the \textit{Alignment} and \textit{Robustness} properties, our method also aims to satisfy the \textit{Contrast} property.
Increasing the distance between dissimilar queries should help distinguish misspelled queries from other distinct queries.
We design the following components for our training method:
\begin{inparaenum}[(i)]
    \item Dual Self-Teaching (DST)
    incorporates the ideas of Dual Learning~\cite{pmlr-v70-xia17a, 10.1145/3471158.3472245} and Self-Teaching~\cite{10.1145/3477495.3531951} to train robust dense retrieval in a bidirectional manner: passage retrieval and query retrieval.
    \item Query Augmentation
    generates a numerous number of misspelling variations for each query to supply our training objective.
\end{inparaenum}

Experimental studies were conducted to assess the efficiency of the proposed method in comparison to existing approaches.
We conduct experiments based on two different pre-trained language models.
We evaluate using two passage retrieval benchmark datasets, a standard one and a specialized one for misspellings robustness evaluation.
For each dataset, we measure performance on both misspelled and non-misspelled queries, where the misspelled queries are both generated and real-world queries.
The experimental results show that the proposed method outperforms the best existing methods for enhancing the robustness of dense retrieval against misspellings without sacrificing performance for non-misspelled queries.

We summarize our contributions as follows:
\begin{compactitem}[•]
    \item We propose a novel training method to enhance the robustness of dense retrieval against misspellings by incorporating three desired properties: \textit{Alignment}, \textit{Robustness}, and \textit{Contrast}.
    \item We introduce Dual Self-Teaching (DST) which adopts the idea of Dual Learning and Self-Teaching to learn robust representations. In addition, we propose Query Augmentation to generate multiple views of a particular query under different misspelling scenarios.
    \item We evaluate our method on misspelled and non-misspelled queries from two passage retrieval datasets. The results show that our method outperforms the previous state-of-the-art methods by a significant margin on misspelled queries.
\end{compactitem}

\begin{figure*}[t]
    \includegraphics[width=12cm, height=5cm]{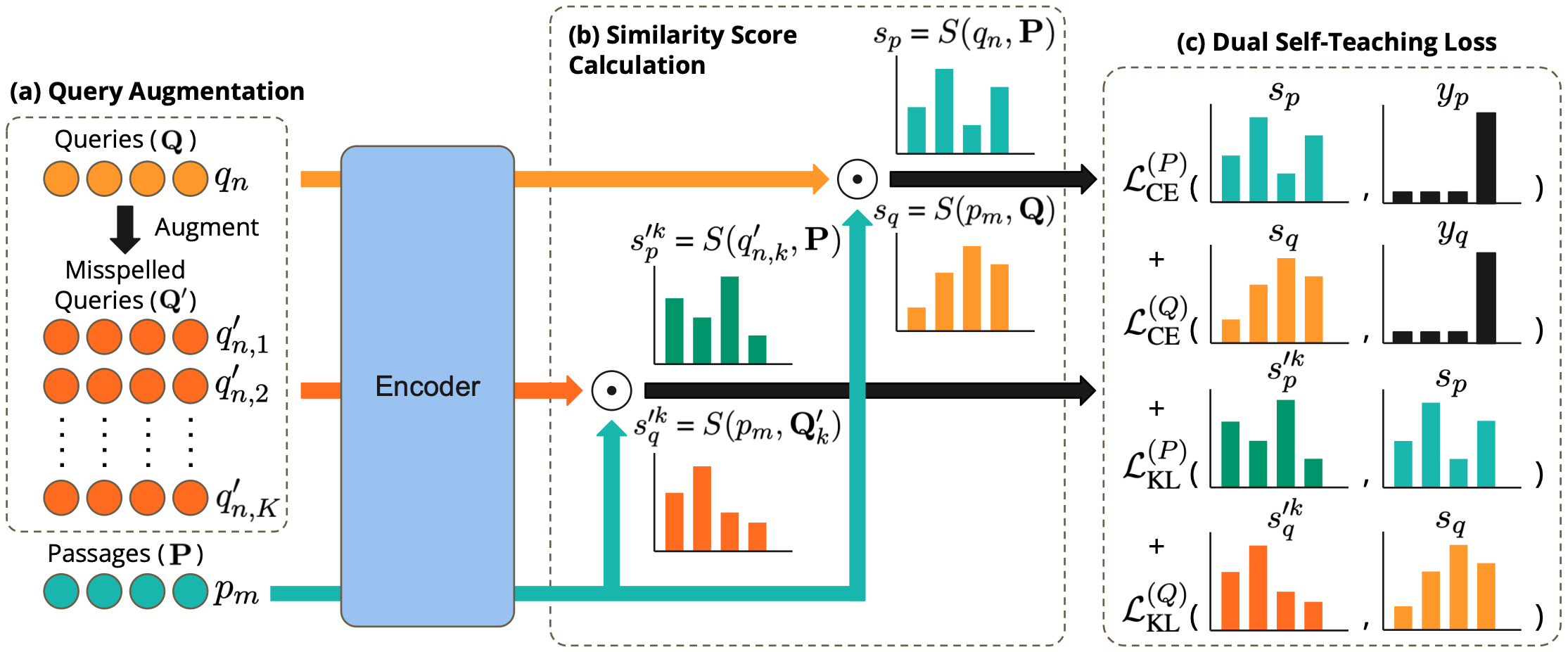}
    \centering
    \caption{The proposed training pipeline consists of three steps: (a) Query Augmentation, (b) Similarity Score Calculation, and (c) Dual Self-Teaching Loss Calculation.}
    \label{fig:1}
\vspace{-5mm}
\end{figure*}

\section{Methodology}
We propose a training pipeline to enhance the dense retrieval capability for handling spelling variations and mistakes in queries. 
As shown in Figure~\ref{fig:1}, the training pipeline comprises three steps.
(i) \emph{Query Augmentation}: we augment each query in the training set into multiple misspelled queries using the typo generators provided by \citet{zhuang-zuccon-2021-dealing}.
(ii) \emph{Similarity Score Calculation}: we compute similarity score distributions between queries and passages for passage retrieval and query retrieval tasks using in-batch negative queries and passages, with additional hard negative passages. 
(iii) \emph{Dual Self-Teaching Loss Calculation}: we compute the DST loss using the similarity score distributions to achieve all three desired properties.

\subsection{Query Augmentation}
The purpose of this step is to guide the learning with a broad array of possible misspelling patterns.
Let $\mathbf{Q}$ denote a set $\{q_1, q_2, ..., q_N\}$ of $N$ queries. 
From all queries in $\mathbf{Q}$,
we generate a set of $K \times N$ misspelled queries $\boldsymbol{\mathcal{Q}}' \text{ = } \{\langle q'_{1,k}, q'_{2,k}, ..., q'_{N,k} \rangle\}_{k=1}^K$, where $K$ is the misspelling variations.
We use five typo generators proposed by \citet{zhuang-zuccon-2021-dealing}, including: RandInsert, RandDelete, RandSub, SwapNeighbor, and SwapAdjacent.
Please refer to Appendix~\ref{appendix:query_augmentation_examples} for examples of the misspelled queries.

\subsection{Similarity Score Calculation}

Let $S(a, \mathbf{B})$ denote a function that computes a similarity score distribution of any vector $a$ over any set of vectors $\mathbf{B}$:

\vspace{-5mm}
\begin{equation} \label{eq:softmax}
    S(a, \mathbf{B}) = \left\{b_i \in \mathbf{B} \,\middle\vert\, \frac{exp(a \cdot b_i)}{\sum_{b_j \in \mathbf{B}} exp(a \cdot b_j)}\right\}
\end{equation}
\vspace{-3mm}


Given $\mathbf{P} \text{ = } \{p_1, p_2, ..., p_M\}$ to be a set of $M$ passages and $\mathbf{Q}'_k \text{ = } \{q'_{1,k}, q'_{2,k}, ..., q'_{N,k}\}$ to be the $k^{th}$ set of misspelled queries in $\boldsymbol{\mathcal{Q'}}$, we compute two groups of score distributions as follows:

\begin{compactitem}[•]
    \item \emph{Passage retrieval}: we calculate score distributions in a query-to-passages direction for each original query $s_p \text{ = } S(q_n, \mathbf{P})$ and misspelled query $s'^k_p \text{ = } S(q'_{n,k}, \mathbf{P})$.
    \item \emph{Query retrieval}: we calculate score distributions in a passage-to-queries direction for original queries $s_q \text{ = } S(p_m, \mathbf{Q})$ and each set of misspelled queries $s'^k_q \text{ = } S(p_m, \mathbf{Q}'_k)$.
\end{compactitem}
This way, we produce four different score distributions ($s_p$, $s'^k_p$, $s_q$, $s'^k_q$) for our training objective.

\subsection{Dual Self-Teaching Loss Calculation}
We design the \textit{Dual Self-Teaching loss} ($\mathcal{L}_\text{DST}$) to capture the three desired properties: \textit{Alignment}, \textit{Robustness}, and \textit{Contrast}.

\vspace{-5mm}
\begin{equation} \label{eq:final_loss}
    \mathcal{L}_\text{DST} = \underbrace{(1 - \beta)\mathcal{L}_\text{DCE}}_{\text{Dual Cross-Entropy}} + \underbrace{\beta\mathcal{L}_\text{DKL}}_{\text{Dual KL-Divergence}}
\end{equation}
\vspace{-3mm}

\textit{Dual Cross-Entropy loss} ($\mathcal{L}_\text{DCE}$) satisfies the \textit{Alignment} and \textit{Contrast} properties by utilizing cross-entropy losses to learn score distributions of the original queries for passage retrieval ($s_p$) and query retrieval ($s_q$) given labels $y_p$ and $y_q$.

\vspace{-5mm}
\begin{equation} \label{eq:dual_ranking}
    \mathcal{L}_\text{DCE} = \underbrace{(1 - \gamma)\mathcal{L}_\text{CE}^{(P)}(s_p, y_p)}_{\text{Passage Retrieval}} + \underbrace{\gamma \mathcal{L}_\text{CE}^{(Q)}(s_q, y_q)}_{\text{Query Retrieval}}
\end{equation}
\vspace{-3mm}

Minimizing the $\mathcal{L}_\text{CE}^{(P)}$ term will increase the similarity scores between queries and their relevant passages to be higher than other irrelevant passages by separating the relevant and irrelevant passages from one another.
Minimizing the $\mathcal{L}_\text{CE}^{(Q)}$ term will increase the similarity scores between passages and their relevant queries to be higher than other irrelevant queries by separating the relevant and irrelevant queries from one another.
In this manner, minimizing one of the two terms will align queries with their corresponding passages, satisfying the \textit{Alignment} property.
Moreover, minimizing both terms will separate queries that refer to different passages and passages that belong to different queries, satisfying the \textit{Contrast} property.

\textit{Dual KL-Divergence loss} ($\mathcal{L}_\text{DKL}$) aims to fulfill the \textit{Robustness} property by using KL losses to match score distributions of misspelled queries $\{s'^1_p, s'^2_p, ..., s'^K_p\}$ and $\{s'^1_q, s'^2_q, ..., s'^K_q\}$ to the score distributions of the original query $s_p$ and $s_q$.

\vspace{-5mm}
\begin{equation} \label{eq:dual_self_distilltion}
    \begin{split}
    \mathcal{L}_\text{DKL} = \frac{1}{K}\sum_{k=1}^K \underbrace{(1 - \sigma)\mathcal{L}_\text{KL}^{(P)}(s'^k_p, s_p)}_{\text{Passage Retrieval Consistency}} \\
    + \underbrace{\sigma \mathcal{L}_\text{KL}^{(Q)}(s'^k_q, s_q)}_{\text{Query Retrieval Consistency}}
    \end{split}
\end{equation}
\vspace{-1mm}

Minimizing $\mathcal{L}_\text{KL}^{(P)}$ and $\mathcal{L}_\text{KL}^{(Q)}$ will reduce the discrepancy between misspelled and non-misspelled queries for both query-to-passages and passage-to-queries score distributions.
This way, we implicitly align representations of the misspelled queries to the original queries, satisfying the \textit{Robustness} property. 
To stabilize training, we apply stop-gradient to the score distributions of the original queries ($s_p$ and $s_q$) in the $\mathcal{L}_\text{DKL}$.
The $\beta$, $\gamma$, and $\sigma$ are the balancing coefficients selected by hyper-parameter tuning on a development set.
With this loss combination, we achieve all three desired properties.

\begin{table*}[h]
    \centering
    \fontsize{8pt}{13pt}
    \selectfont
    \scalebox{0.92}{
        \makebox[\linewidth]{
            \tabcolsep=0.12cm
            \begin{tabular}{|l|p{0.4cm} l p{0.4cm} l|p{0.4cm} l p{0.4cm} l p{0.4cm} l|p{0.4cm} l p{0.4cm} l|p{0.4cm} l p{0.4cm} l p{0.4cm} l|}
                \hline
                & \multicolumn{10}{c|}{BERT-based} & \multicolumn{10}{c|}{CharacterBERT-based} \\
                \cline{2-21}
                & \multicolumn{4}{c|}{MS MARCO} & \multicolumn{6}{c|}{DL-typo} & \multicolumn{4}{c|}{MS MARCO} &  \multicolumn{6}{c|}{DL-typo} \\
                Methods & \multicolumn{2}{c}{MRR@10} & \multicolumn{2}{c|}{R@1000} & \multicolumn{2}{c}{nDCG@10} & \multicolumn{2}{c}{MRR} & \multicolumn{2}{c|}{MAP} & \multicolumn{2}{c}{MRR@10} & \multicolumn{2}{c|}{R@1000} & \multicolumn{2}{c}{nDCG@10} & \multicolumn{2}{c}{MRR} & \multicolumn{2}{c|}{MAP} \\
                \hline
                DPR & .143 & (.331) & .696 & (\textbf{.954}) & .276 & (\textbf{.682}) & .431 & (\textbf{.873}) & .175 & (.563) & .162 & (.321) & .726 & (.945) & .268 & (.643) & .376 & (\underline{.832}) & .212 & (.503) \\
                + Aug & .227 & (.334) & .857 & (.950) & \underline{.398} & (\textbf{.682}) & .530 & (.806) & \underline{.286} & (\underline{.565}) & .258 & (.326) & .883 & (.946) & .414 & (.631) & .578 & (.783) & .318 & (.512) \\
                + Aug + CL & .234 & (\underline{.335}) & .867 & (\underline{.951}) & .387 & (.668) & \underline{.536} & (\underline{.864}) & .267 & (.544) & .263 & (\underline{.330}) & .894 & (\underline{.947}) & .466 & (\textbf{.677}) & \underline{.635} & (.819) & \underline{.360} & (\textbf{.544}) \\
                + ST & \underline{.237} & (.333) & \underline{.874} & (.950) & .392 & (\underline{.677}) & .525 & (.852) & .283 & (.557) & \underline{.274} & (\textbf{.332}) & \underline{.900} & (\underline{.947}) & \underline{.469} & (.650) & .619 & (.810) & .359 & (.517) \\
                \hline
                + DST (our) & \textbf{.260}$\dagger$ & (\textbf{.336}) & \textbf{.894}$\dagger$ & (\textbf{.954}) & \textbf{.432} & (.673) & \textbf{.558} & (.833) & \textbf{.343}$\dagger$ & (\textbf{.568}) & \textbf{.288}$\dagger$ & (\textbf{.332}) & \textbf{.918}$\dagger$ & (\textbf{.949}) & \textbf{.529}$\dagger$ & (\underline{.673}) & \textbf{.742}$\dagger$ & (\textbf{.854}) & \textbf{.403} & (\underline{.537}) \\
                \hline
            \end{tabular}
        }
    }
    \caption{\label{font-table} Results of different training methods on misspelled and non-misspelled queries. We report the results in the format of {\tt "misspelled query performance (non-misspelled query performance)"}. We emphasize the best score with bold text and the second-best score with underlined text. We use $\dagger$ to denote DST results that significantly outperform the second-best result ($p<0.05$).}
\label{tab:main_results}
\vspace{-3mm}
\end{table*}

\section{Experimental Settings}

\subsection{Training Details} 
We experiment on two pre-trained language models, BERT~\cite{devlin-etal-2019-bert} and CharacterBERT~\cite{el-boukkouri-etal-2020-characterbert}.
We train models only on the training set of MS MARCO dataset~\cite{DBLP:conf/nips/NguyenRSGTMD16}.
Moreover, the training data provided by the Tevatron toolkit~\cite{Gao2022TevatronAE} also contains hard negative passages.
We include the training set details and hyper-parameter settings in Appendix~\ref{appendix:training_setting}.

\subsection{Competitive Methods}
To show the effectiveness of our method, we compare our work with the following baseline and competitive training methods.
\begin{compactitem}[•]
    \item \textit{DPR}~\cite{karpukhin-etal-2020-dense} is a baseline training method that trains dense retrieval merely on non-misspelled queries using $\mathcal{L}_\text{CE}^{(P)}$ loss.
    \item \textit{DPR+Aug}~\cite{zhuang-zuccon-2021-dealing} is the Typos-aware Training method which trains dense retrieval on both misspelled and non-misspelled queries using $\mathcal{L}_\text{CE}^{(P)}$ loss.
    \item \textit{DPR+Aug+CL}~\cite{Sidiropoulos_2022} employs additional contrastive loss to train the misspelled queries. 
    \item \textit{DPR+ST}~\cite{10.1145/3477495.3531951} is the Self-Teaching method that trains dense retrieval on both misspelled and non-misspelled queries using $\mathcal{L}_\text{CE}^{(P)}$ and $\mathcal{L}_\text{KL}^{(P)}$ losses.
\end{compactitem}
Note that their query augmentation method is identical to the Query Augmentation with $K \text{ = } 1$.
We retrain all models using the same setting described in the previous section.

\subsection{Dataset and Evaluation}
\textbf{Datasets.}
We evaluate the effectiveness of DST on two passage retrieval datasets, MS MARCO and DL-typo~\cite{10.1145/3477495.3531951}, each with misspelled and non-misspelled queries.
There are 8.8 million candidate passages for both datasets.
The development set of MS MARCO contains 6,980 non-misspelled queries.
To obtain misspelled queries, we use the typos generator method proposed by \citet{zhuang-zuccon-2021-dealing} to generate 10 misspelled variations for each original query.
The DL-typo provides 60 real-world misspelled queries and 60 corresponding non-misspelled queries that are corrected manually.

\noindent
\textbf{Evaluation.}
We use the standard metrics originally used by each dataset's creators.
For MS MARCO, each misspelled query performance is the average of 10 measurements.
We employ Ranx evaluation library~\cite{bassani2022ranx} to measure performance and statistical significance.
Specifically, we use a two-tailed paired t-test with Bonferroni correction to measure the statistical significance $(p<0.05)$.

\section{Experimental Results}

\subsection{Main Results}
As shown in Table~\ref{tab:main_results}, the results indicate that DST outperforms competitive methods for misspelled queries in every case without sacrificing performance for non-misspelled queries in eight out of ten cases.
We observe some performance trade-offs for the BERT-based model in the DL-typo dataset's non-misspelling scores (nDCG@10 and MRR).
Aside from that, there is no performance trade-off for the CharacterBERT-based model.
These outcomes conform with the observation in Figure~\ref{fig:query_distribution} (Section~\ref{experiment:query_representations_distribution}) that DST improves the \textit{Robustness} and \textit{Contrast} of misspelled queries.

\subsection{Query Augmentation Size Study}
To study the benefit of query augmentation and find the optimal augmentation size, we measure the performance of BERT-based dense retrieval models trained with DST using the query augmentation size $K$ of 1, 10, 20, 40, and 60.
Note that the query augmentation method used in previous works is a special case of Query Augmentation when $K\text{ = }1$.
We report the results using MRR@10 for the development set of the MS MARCO dataset.
We also report training time to show trade-offs between performance and computation.

\begin{table}[h]
    \fontsize{8pt}{13pt}
    \selectfont
    \makebox[\linewidth]{
        \begin{tabular}{|l|c c c c c|}
            \hline
            \multirow{2}{*}{Queries} & \multicolumn{5}{|c|}{$K$} \\
            & 1 & 10 & 20 & 40 & 60 \\
            \hline
            Original & .334 & .334 & \underline{.335} & \textbf{.336} & .332 \\
            Misspelled & .251 & \underline{.258} & \textbf{.260} & \textbf{.260} & \textbf{.260} \\
            \hline
            Training time (hr) & 18 & 20 & 23 & 31 & 39 \\
            \hline
        \end{tabular}
    }
    \caption{\label{font-table} Results of query augmentation size study. We train all models in this experiment on a V100 32G GPU.}
\label{tab:augmentation_size}
\end{table}

As shown in Table~\ref{tab:augmentation_size}, the results indicate that increasing $K$ improves the performance of both misspelled and non-misspelled queries, but only up to a certain point, after which the performance begins to decline.
We observe that setting $K\text{ = }40$ produces the best results, and there is no further performance improvement after this point.
%

\subsection{Loss Ablation Study}
In this experiment, we study the benefit of each term in DST by training BERT-based dense retrieval models on variant loss combinations with $K \text{ = } 40$.
\begin{table}[h]
    \fontsize{8pt}{13pt}
    \selectfont
    \scalebox{0.92}{
    \makebox[\linewidth]{
        \begin{tabular}{|c c c c|c|}
            \hline
            $\mathcal{L}_\text{CE}^{(P)}$ & $\mathcal{L}_\text{CE}^{(Q)}$ & $\mathcal{L}_\text{KL}^{(P)}$ & $\mathcal{L}_\text{KL}^{(Q)}$ & MRR@10 \\
            \hline
            \checkmark & \checkmark & \checkmark & \checkmark & \textbf{.260} (\underline{.336}) \\
            \hline
            \checkmark & \checkmark & \checkmark & & \underline{.257} (.335) \\
            \checkmark & \checkmark & & \checkmark & .228 (.326) \\
            \checkmark & & \checkmark & \checkmark & .251 (\textbf{.337}) \\
            & \checkmark & \checkmark & \checkmark & .087 (.114) \\
            \hline
            \checkmark & & \checkmark & & .249 (\underline{.336}) \\
            & \checkmark & & \checkmark & .120 (.158) \\
            \hline
        \end{tabular}}
    }
    \caption{\label{font-table} Loss ablation study results on MS MARCO.}
\label{tab:ablation_study}
\vspace{-3mm}
\end{table}
%
\noindent
The results in Table~\ref{tab:ablation_study} reveal that $\mathcal{L}_\text{KL}^{(P)}$ and $\mathcal{L}_\text{KL}^{(Q)}$ terms positively contribute to the performance of misspelled and non-misspelled queries, with the $\mathcal{L}_\text{KL}^{(P)}$ being more significant.
The $\mathcal{L}_\text{CE}^{(P)}$ term is crucial for retrieval performance, whereas the $\mathcal{L}_\text{CE}^{(Q)}$ term indirectly improves the performance of misspelled queries by separating their pristine queries from the surrounding queries.
Disabling query retrieval terms ($\mathcal{L}_\text{CE}^{(Q)}$ and $\mathcal{L}_\text{KL}^{(Q)}$) greatly reduces performances for misspelled queries.
The passage retrieval terms ($\mathcal{L}_\text{CE}^{(P)}$ and $\mathcal{L}_\text{KL}^{(P)}$) are indispensable and cannot be substituted.

\subsection{Query Distributions} \label{experiment:query_representations_distribution}
\begin{figure}[ht!]
\vspace{-3mm}
     \centering
     \begin{subfigure}[b]{0.5\textwidth}
         \centering
         \includegraphics[width=\textwidth]{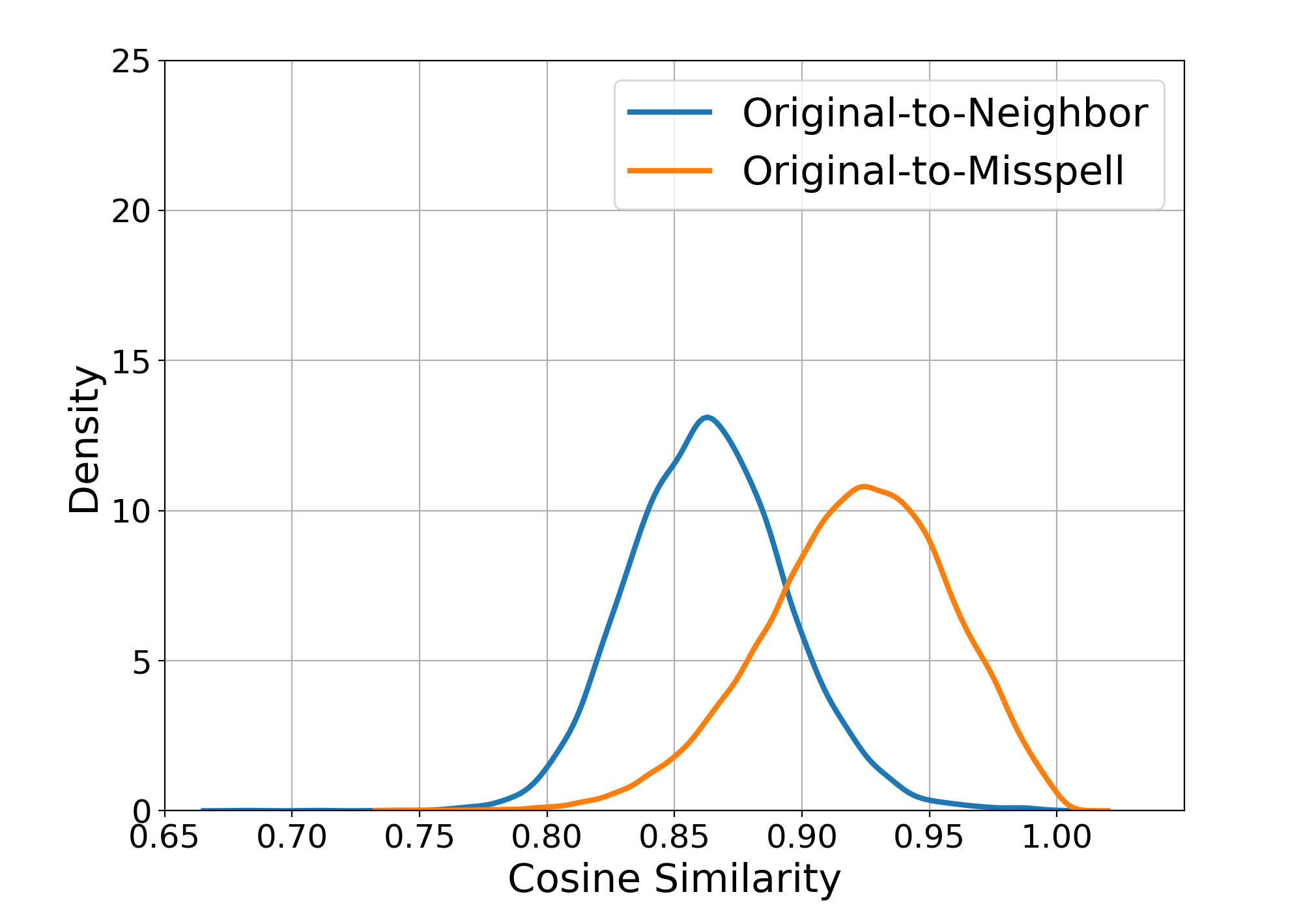}
         \caption{DPR~\cite{karpukhin-etal-2020-dense}.}
         \label{fig:dpr}
     \end{subfigure}
     \hfill
     \begin{subfigure}[b]{0.5\textwidth}
         \centering
         \includegraphics[width=\textwidth]{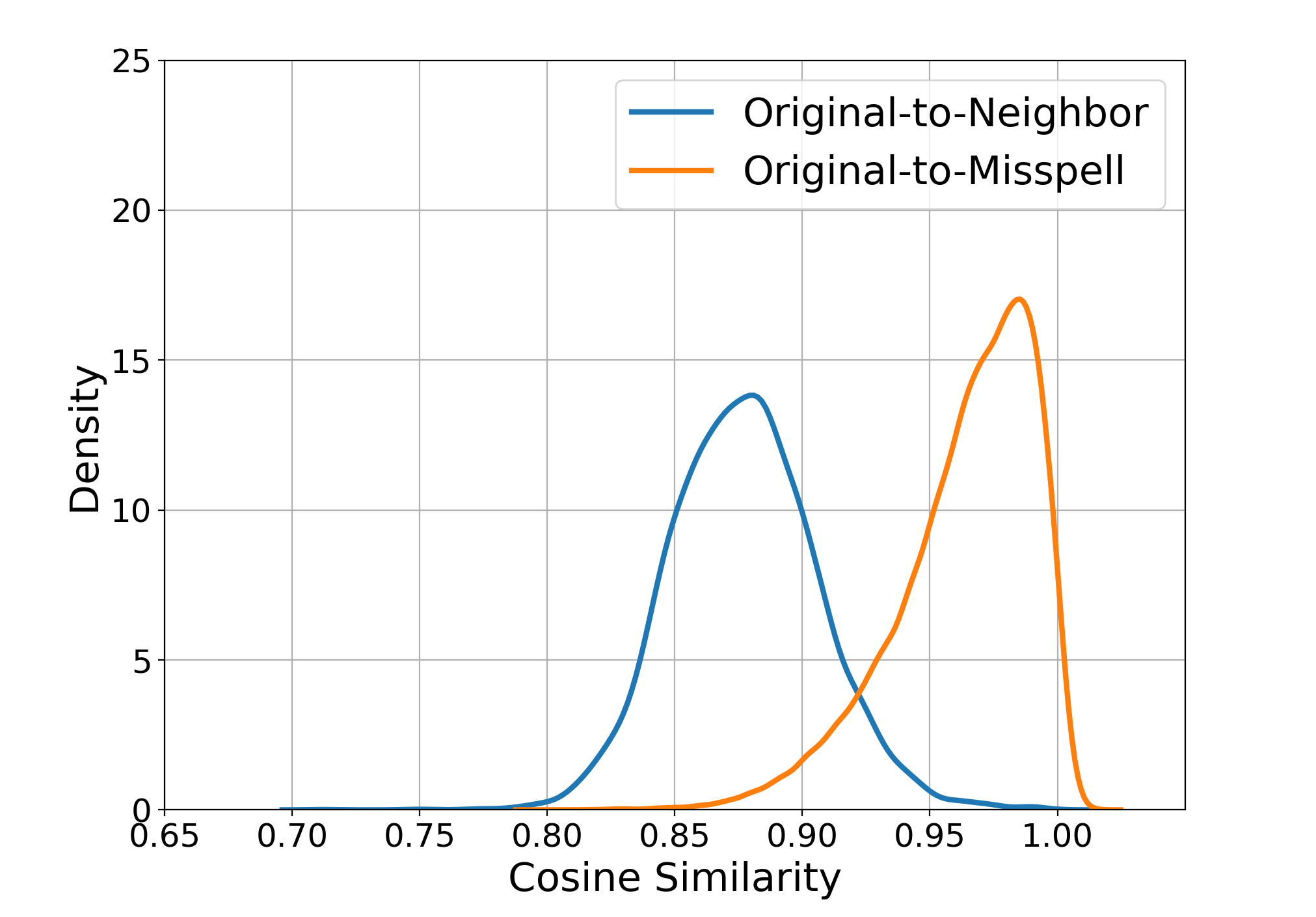}
         \caption{Self-Teaching~\cite{10.1145/3477495.3531951}.}
         \label{fig:dpr_st}
     \end{subfigure}
     \hfill
     \begin{subfigure}[b]{0.5\textwidth}
         \centering
         \includegraphics[width=\textwidth]{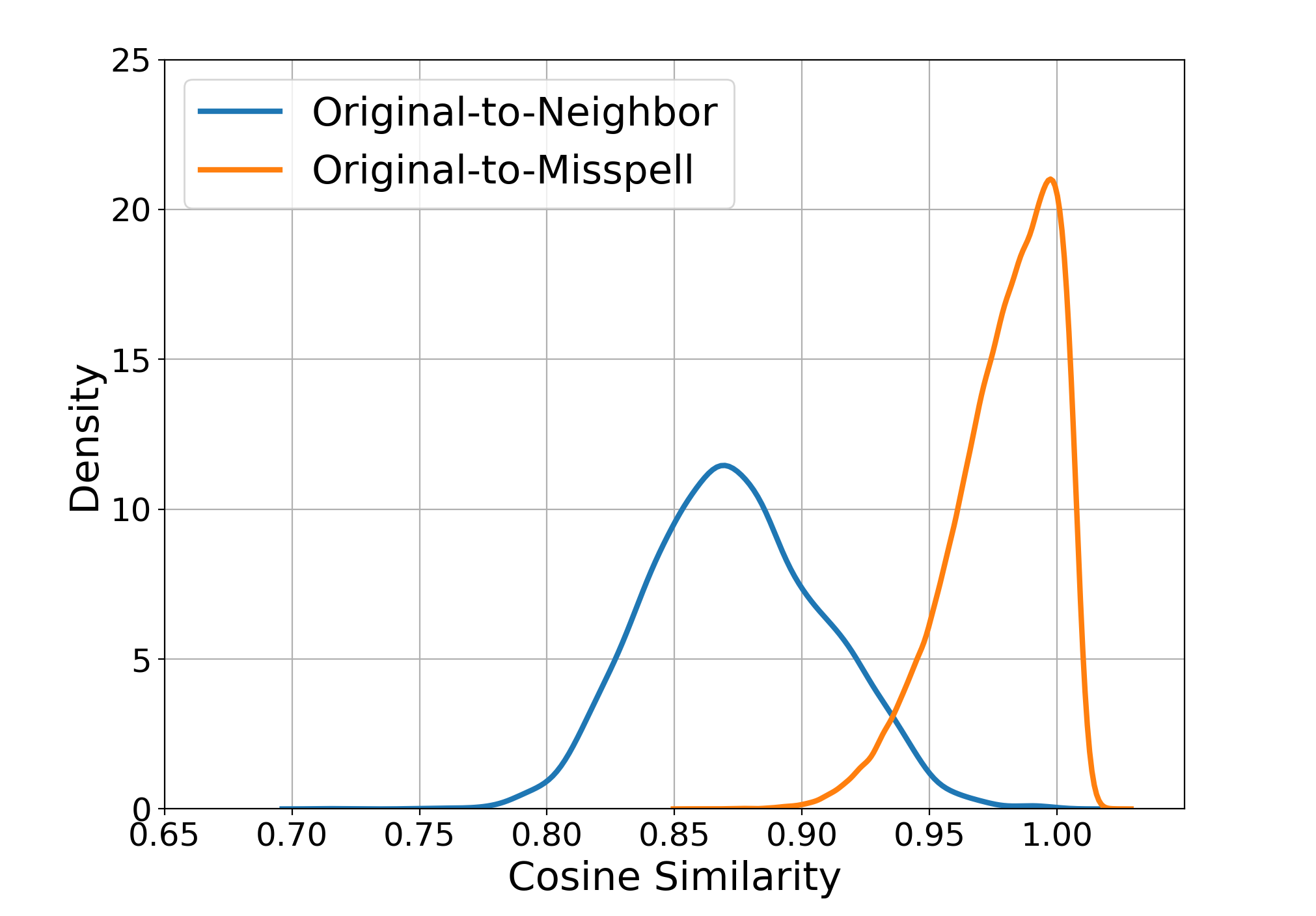}
         \caption{Dual Self-Teaching (our).}
         \label{fig:dpr_dst}
     \end{subfigure}
        \caption{Kernel density of Original-to-Neighbor (orange) and Original-to-Misspell (blue) of different training methods.}
        \label{fig:query_distribution}
    \label{fig:distribution}
\vspace{-5mm}
\end{figure}

The purpose of this section is to study the impact of our training method on the \textit{Robustness} and \textit{Contrast} of misspelled queries.
We also compare our method against the baseline and competitive methods to show its effectiveness.
The \textit{Robustness} and \textit{Contrast} of misspelled queries are illustrated using the following kernel density graphs:
\begin{compactitem}[•]
    \item Original-to-Misspell: the cosine similarity distribution between original and misspelled queries.
    \item Original-to-Neighbor: the cosine similarity distribution between original and neighbor queries.
\end{compactitem}
The \textit{Robustness} property is emphasized by the Original-to-Misspell distribution having high cosine similarity.
On the other hand, the \textit{Contrast} property is emphasized by the small overlapping between Original-to-Misspell and Original-to-Neighbor distributions.
The results in Figure~\ref{fig:distribution} show that our method (c) produces the best \textit{Robustness} and \textit{Contrast} properties for misspelled queries in comparison to other methods.

\section{Conclusion}
This paper aims to address the misspelling problem in dense retrieval.
We formulate three desired properties for making dense retrieval robust to misspellings: \textit{Alignment}, \textit{Robustness}, and \textit{Contrast}.
Unlike previous methods, which only focus on the \textit{Alignment} and \textit{Robustness} properties, our method considers all the desired properties.
The empirical results show that our method performs best against misspelled queries, revealing the importance of the \textit{Contrast} property for handling misspellings.

\section{Limitations}
We list the limitations of our work as follows:
\begin{compactitem}[•]
    \item The Query Augmentation is designed for the English alphabet; therefore, other languages with different alphabets will require further work.
    \item Since the training strategy relies on fine-tuning a pre-trained language model using a large passage retrieval dataset, it may not be suitable for languages with limited resources
\end{compactitem}

\bibliography{anthology,custom}
\bibliographystyle{acl_natbib}

\clearpage
\appendix

\section{Appendix}

\subsection{Training Setup and Hyperparameters} \label{appendix:training_setting}
The MS MARCO is a large-scale English language dataset for machine reading comprehension (MRC).
The dataset consists of anonymized queries sampled from Bing's search query logs, each with human generated answers.
The training set we used contains 400,782 training samples, each consisting of a query, positive passage, and a set of hard negative passages, which we randomly select 7 hard negative passages for each training sample.
We set a batch size to 16 and use in-batch negative sampling for each training sample. Therefore, we obtain 7 + 8 * 15 = 127 negative passages for each training sample.
We use the AdamW optimizer and learning rate of $1\mathrm{e-}5$ for 150,000 steps with a linear learning rate warm-up over the first 10,000 steps and a linear learning rate decay over the rest of the training steps.
For our training method, we set the hyper-parameters $\beta\text{ = }0.5$, $\gamma\text{ = }0.5$, $\sigma\text{ = }0.2$, and the query augmentation size $K\text{ = }40$.
Using one V100 32G GPU, the BERT-based model training time is around 31 hours, while the CharacterBERT-based model training time is roughly 56 hours.

\subsection{Query Augmentation Examples} \label{appendix:query_augmentation_examples}
Table \ref{tab:misspelled_queries_examples} provides examples of misspelled queries generated by the Query Augmentation for each original query.

\begin{table}[h]
    \fontsize{8pt}{13pt}
    \selectfont
    \makebox[\linewidth]{
        \begin{tabular}{|l|}
            \hline
            \textbf{Original query:} \\
            what is the goddess of agriculture in greek mythology \\
            \textbf{Misspelled queries:} \\
            what is the \colorbox{blue!30}{goddoess} of agriculture in greek mythology \\
            what is the goddess of \colorbox{red!30}{agriulture} in greek mythology \\
            what is the goddess of agriculture in greek \colorbox{orange!30}{mythologo} \\
            what is the \colorbox{yellow!30}{goddses} of agriculture in greek mythology \\
            what is the goddess of agriculture in greek \colorbox{green!30}{myhhology} \\
            what is the goddess of agriculture in \colorbox{blue!30}{greeck} mythology \\
            what is the goddess of agriculture in greek \colorbox{red!30}{myhology} \\
            what is the goddess of agriculture in \colorbox{orange!30}{grvek} mythology \\
            what is the goddess of \colorbox{yellow!30}{agricultrue} in greek mythology \\
            what is the goddess of \colorbox{green!30}{ahriculture} in greek mythology \\
            \hline
        \end{tabular}
    }
    \caption{\label{font-table} The outputs of Query Augmentation with $K\text{ = } 10$. We use different colors to indicate different types of typo: \colorbox{blue!30}{RandInsert}, \colorbox{red!30}{RandDelete}, \colorbox{orange!30}{RandSub}, \colorbox{yellow!30}{SwapNeighbor}, and \colorbox{green!30}{SwapAdjacent}.}
\label{tab:misspelled_queries_examples}
\end{table}

\subsection{Licenses}
\textbf{Datasets}: The MS MARCO dataset is available under the MIT license, and the DL-typo dataset is available under the Apache license 2.0.
These licenses allow users to use the datasets under non-restrictive agreements.

\noindent
\textbf{Softwares}: We employ Hugging Face~\cite{wolf-etal-2020-transformers} and Tevatron~\cite{Gao2022TevatronAE} libraries to train dense retrieval models.
We utilize Ranx library~\cite{bassani2022ranx} to evaluate retrieval performance.
These libraries are available under the Apache license 2.0 which allows both academic and commercial usages.
For this reason, we release our code under the Apache license 2.0 to make our code fully accessible and compatible with the other codes we use.

\end{document}